\documentclass[acmsmall,screen]{acmart}

\AtBeginDocument{%
  }

\usepackage{amsmath,amsfonts}
\usepackage{algorithmic}
\usepackage{graphicx}
\usepackage{textcomp}
\usepackage[ruled,vlined,linesnumbered]{algorithm2e}
\usepackage{multirow}
\usepackage{adjustbox}
\usepackage{pifont}
\usepackage{threeparttable}
\usepackage{booktabs}
\usepackage{bbding}
\usepackage{tablefootnote}
\usepackage{hyperref}
\usepackage{tcolorbox}
\usepackage{enumitem}
\usepackage{enumerate}
\usepackage{subfig}
\usepackage{url}
\usepackage{soul}
\usepackage{balance}
\usepackage{makecell}
\hypersetup{
colorlinks=true,
urlcolor=black,
linkcolor=black,
citecolor=black
}

\usepackage{ulem}
\usepackage{collcell}
\usepackage[strict]{changepage}
  \definecolor{ABlue}{HTML}{127bca}
 \definecolor{LHScolor}{HTML}{555555}
\usepackage{framed}
\definecolor{formalshade}{rgb}{1.0,1.0,1.0}
\definecolor{side}{rgb}{0.0,0.2,0.6}

\newcommand\tian[1]{\textcolor{black!90}{#1}}

\newcommand{\fpnum}{497}

\setcopyright{cc}
\setcctype{by-nc-nd}
\acmJournal{PACMSE}
\acmYear{2026} \acmVolume{3} \acmNumber{FSE} \acmArticle{FSE054}
\acmMonth{7} \acmDOI{10.1145/3797082}

\begin{document}

\title{Characterizing and Mitigating False-Positive Bug Reports in the Linux Kernel}

\author{Jiashuo Tian}
\orcid{0009-0008-4043-3675}
\affiliation{%
  \institution{School of Computer Software, Tianjin University}
  \city{Tianjin}
  \country{China}
}
\email{tianjiashuo@tju.edu.cn}

\author{Dong Wang}
\authornote{Corresponding author (w\_dong@tju.edu.cn), and contributed equally to Jiashuo Tian}
\orcid{0000-0002-2004-0902}
\affiliation{%
  \institution{School of Computer Software, Tianjin University}
  \city{Tianjin}
  \country{China}
}
\email{dong_w@tju.edu.cn}

\author{Chen Yang}
\orcid{0000-0003-0759-940X}
\affiliation{%
  \institution{School of Computer Software, Tianjin University}
  \city{Tianjin}
  \country{China}
}
\email{yangchenyc@tju.edu.cn}

\author{Haichi Wang}
\orcid{0009-0007-6953-8369}
\affiliation{%
  \institution{School of Computer Software, Tianjin University}
  \city{Tianjin}
  \country{China}
}
\email{wanghaichi@tju.edu.cn}

\author{Zan Wang}
\orcid{0000-0001-6173-8170}
\affiliation{%
  \institution{School of Computer Software, School of Artificial Intelligence, Tianjin University}
  \city{Tianjin}
  \country{China}
}
\email{wangzan@tju.edu.cn}

\author{Junjie Chen}
\orcid{0000-0003-3056-9962}
\affiliation{%
  \institution{School of Computer Software, Tianjin University}
  \city{Tianjin}
  \country{China}
}
\email{junjiechen@tju.edu.cn}

\begin{abstract}

False-positive bug reports represent a significant yet underexplored challenge in the development and maintenance of the Linux kernel. 
They occur when correct system behavior is mistakenly flagged as a defect, consuming developer effort without leading to actual code improvements. Such reports can mislead developers, waste debugging resources, and delay the resolution of real bugs.
In this paper, we present the first comprehensive empirical study of false-positive bug reports in the Linux kernel. 
We manually construct a dataset of 2,006 bug reports comprising 1,509 genuine bugs and 497 false positives collected from Bugzilla and Syzkaller. 
Our analysis indicates that false positives demand effort comparable to real bugs, often requiring extended discussions and non-trivial closure time. 
They occur in several components, especially File Systems and Drivers, mainly due to external dependencies and semantic misunderstandings. 
To address this challenge, we evaluate large language models (LLMs) for automated false-positive bug report mitigation. 
Among various prompting strategies, retrieval-augmented generation (RAG) performs best, achieving 91\% recall and an F1 score of 88\%.
These findings highlight the non-negligible cost of false positive bug reports and show the promise of LLMs for more efficient false positive mitigation in the Linux kernel.

\end{abstract}



\maketitle

\section{Introduction}
\label{sec:intro}

The Linux kernel is one of the most widely deployed software systems, serving as the core of numerous operating systems that power servers, desktops, mobile devices, and cloud services~\cite{mu2022depth,tan2023syzdirect}. Given its central role in the software ecosystem, ensuring the reliability and correctness of the Linux kernel is essential. Even a single bug can trigger severe consequences, ranging from system crashes to critical security vulnerabilities. However, maintaining such reliability is challenging. Diagnosing and fixing kernel bugs is particularly difficult because the codebase is massive and evolving, closely tied to diverse hardware, and heavily reliant on concurrency and synchronization, which together demand specialized expertise from experienced maintainers~\cite{mu2022depth}.

The challenge of debugging the Linux kernel is further complicated by the prevalence of \textit{false-positive bug reports}. Bug tracking systems such as Bugzilla~\cite{bugzilla} and Syzkaller~\cite{syz} continuously generate reports of potential bugs. While many of these reports correspond to genuine issues, a significant fraction are false-positive reports that do not indicate actual kernel defects. These may arise from misconfigured environments, incorrect assumptions, or misinterpretations of expected kernel behaviors. 
For example, a user reported that KernelShark failed to display graphs correctly,\footnote{\url{https://bugzilla.kernel.org/show_bug.cgi?id=217460}} which initially appeared as a kernel visualization bug but was ultimately traced to a missing \textit{`fonts-ttf-freefont'} dependency. False-positive bug reports can consume valuable developer resources, since maintainers must still analyze and attempt to reproduce the reported issue before concluding that no genuine bug exists. Prior studies indicate that a substantial proportion of invalid or low-quality bug reports in large-scale projects can often lead to frustration among developers and delays in addressing true defects~\cite{he2020deep, laiq2024industrial, laiq2023data, d3}. In the Linux ecosystem, where timely responses to bugs are critical for both stability and security, the cost of triaging false positives is further amplified.

Although false-positive bug reports in the Linux kernel pose a significant practical challenge, research on this phenomenon remains largely unexplored. 
Prior empirical studies have largely examined the quality of bug reports in traditional software projects such as Eclipse~\cite{eclipse}, Mozilla~\cite{mozilla}, and Apache~\cite{apache},
where invalid reports are typically classified as false positives, non-reproducible cases, or those lacking sufficient details~\cite{anvik2005coping, laiq2023data, sun2011bug}. 
These studies attribute invalid reports to user mistakes, misunderstandings of expected functionality, or issues from external libraries.
However, the Linux kernel differs markedly from these projects, limiting the applicability of such findings.
First, it is exceptionally large and complex, involving extensive concurrency, synchronization, and low-level memory management. 
Second, kernel development is intertwined with diverse hardware platforms and device drivers, leading to hardware-related misleading reports.
Third, unlike most user-facing applications, a substantial portion of kernel bug reports are generated by automated fuzzing tools such as Syzkaller, which produce large volumes of reports with intricate execution traces that demand specialized tools and deep kernel expertise.
Collectively, these distinctions highlight that existing taxonomies and solutions cannot be directly applied to the Linux kernel.

To address this gap, we conduct the first empirical study of false-positive bug reports in the Linux kernel to facilitate the understanding of their characteristics and explore potential mitigation strategies. 
To enable a comprehensive evaluation, 
we collected 2,006 bug reports from two major Linux bug trackers (i.e., Bugzilla and Syzkaller), among which \fpnum{} were manually labeled as false-positive bug reports (accounting for 24.78\%).
This underscores their significant presence in Linux kernel development.
We then empirically investigate three research questions:

\begin{itemize}

\item \textbf{RQ1: To what extent do false-positive bug reports in the Linux kernel consume developer time and effort?}
This RQ investigates the practical burden that false positives impose on developers. We measure indicators such as the length of discussions, the number of participants, and the time to resolution in order to assess whether handling false positives demands a level of effort comparable to resolving genuine bugs.

\item \textbf{RQ2: How are false-positive bug reports distributed across different Linux kernel components?}
The Linux kernel is organized into several major components (e.g., File System, Drivers, Networking, etc.), each with distinct roles, dependencies, and failure modes. This RQ examines how false positives are spread across these components, aiming to identify which areas of the kernel are particularly susceptible to spurious reports.

\item \textbf{RQ3: What are the root causes of false-positive bug reports in the Linux kernel?}
This RQ aims to identify and categorize the underlying reasons for false-positive bug reports. 
A clear understanding of these causes is essential for guiding the development of mitigation strategies that both reduce wasted engineering effort and enhance confidence in automated validation pipelines.

\end{itemize}

Based on our empirical analysis, we derive the following key findings.
For \textbf{RQ1}, our results show that false-positive bug reports are non-trivial to handle and require substantial developer effort comparable to that of genuine bugs, with average closure times reaching 224.08 days on Bugzilla and 120.52 days on Syzkaller. This suggests that false positives represent a non-negligible source of hidden cost in kernel maintenance and strengthen our motivation to minimize the effort involved in mitigating false-positive bug reports.
For \textbf{RQ2}, we find that false positives cluster in certain major kernel components. The Drivers and File System contribute to the highest proportions (23.42\% and 21.15\%), together accounting for over 40\% of all false-positive bug reports, suggesting the need for better documentation and better validation of edge-case behaviors and hardware interactions.
For \textbf{RQ3}, false-positive reports were classified into four main reasons and ten sub-reasons. The majority stem from two primary root causes: External Dependency Issues (46.08\%) and Misunderstandings of Features or Limitations (39.24\%).
Furthermore, our stage-wise correlation analysis reveals that each component of the Linux kernel is typically associated with specific dominant root causes. 

We next explore whether automated techniques can mitigate false-positive bug reports in the Linux kernel. 
Existing work on detecting invalid reports~\cite{anvik2005coping, laiq2023data, sun2011bug, d3, blade} has mainly used datasets from conventional software projects and targeted categories such as duplicates or incomplete submissions, making them unsuitable for the kernel’s unique failure patterns and debugging practices. 
In addition, existing classifiers typically operate as opaque black boxes with limited explainability, further hindering their practical adoption.
Motivated by this gap, we formulate our final research question.

\begin{itemize}

\item \textbf{RQ4: Can large language models (LLMs) help mitigate false-positive bug reports in the Linux kernel?} 
Given their recent success in code understanding and automated debugging, LLMs offer a promising avenue for identifying and explaining false positives. 
Building on stage and root-cause analyses, we investigate the extent to which LLMs can accurately classify bug reports as genuine bugs or false positives while providing meaningful interpretations of their decisions.

\end{itemize}

\noindent
Our results demonstrate that LLMs can effectively assist in identifying false-positive bug reports, though their performance varies by prompting strategy. Zero-shot achieves only moderate accuracy, while enhanced zero-shot improves substantially, and retrieval-augmented generation (RAG) yields the best results (Accuracy = 0.81, Recall = 0.91, F1 = 0.88). Moreover, LLMs provide generally reliable explanations (85.29\% judged accurate), indicating their potential as both classifiers and interpretable assistants for Linux kernel triage.

\smallskip
\noindent
\textbf{Contributions}. In summary, this work makes the following key contributions: (1) We present the first comprehensive empirical investigation of false-positive bug reports in the Linux kernel by characterizing their prevalence, developer cost, distribution across kernel components, and root causes.
(2) Through diverse prompting strategies, we assess the ability of LLMs to identify and interpret false positives in the Linux kernel, demonstrating their strong potential.
(3) We provide actionable recommendations for both researchers and practitioners based on empirical findings. To facilitate future research, we have released a public replication package~\cite{replication} including the manually annotated Linux kernel false-positive bug reports, along with their metadata.

\section{Background and Related Work}
\label{sec:background}

\subsection{The Linux Kernel and Its Main Components}
The Linux kernel is the central component of modern operating systems, acting as the intermediary between hardware and userspace applications. It is responsible for process scheduling, memory management, I/O coordination, and enforcing security policies, ensuring that applications can run efficiently and safely on diverse hardware platforms. Unlike traditional software systems, the kernel must maintain compatibility with an enormous spectrum of devices and environments, from embedded systems and smartphones to high-performance servers and cloud infrastructures. This breadth of deployment contributes to its scale (over thirty million lines of code maintained by thousands of contributors worldwide) and to the inherent complexity of its architecture.

To manage such complexity, the kernel is organized into several major components, each responsible for distinct aspects of system operation~\cite{bugzilla_cgi}:
(1) The \textbf{Drivers} component connects the kernel to peripheral hardware, encompassing a vast collection of device-specific modules. 
(2) The \textbf{File System} component provides abstractions for files and directories while supporting a variety of file system formats such as ext4, Btrfs, and XFS. 
(3) The \textbf{Tools} component aids in profiling, debugging, and performance analysis. 
(4) The \textbf{Networking} component implements protocol stacks and packet processing pipelines, enabling communication across machines and networks. 
(5) At the heart of the system lies the \textbf{Kernel Core}, which manages essential services such as CPU scheduling, synchronization primitives, and memory allocation. 
(6) The \textbf{Security} component enforces isolation and access control policies through mechanisms such as SELinux and AppArmor. 
(7) The \textbf{IO} component manages input/output subsystems such as block I/O, character devices, and storage management.
While the kernel contains many other components, our study finds that these additional components contribute only a very small fraction of false-positive reports. 
Since our study focuses on false-positive bug reports, we omit their detailed introduction here  for brevity, a full introduction can be found on our project homepage~\cite{replication}. 
Together, these components illustrate the functional diversity and technical intricacy of the Linux kernel. They also provide the structural context for understanding how false-positive bug reports are distributed across different parts of the kernel. Each component presents unique challenges for bug detection and diagnosis, and therefore differences in the volume, characteristics, and root causes of false positives are to be expected~\cite{abal201442,tan2023syzdirect}.

\subsection{Related Work}
\label{sec:related}

\textbf{Empirical Studies of Kernel Bugs.}
A number of studies have conducted empirical investigations into kernel bugs, though with different goals than ours. Abal et al.~\cite{abal201442} examined 42 Linux kernel bugs and found that variability bugs are not tied to specific bug types, error-prone features, or code locations, but that variability itself significantly increases bug complexity. PDiff~\cite{jiang2020pdiff} provided a systematic study of the patch presence testing problem, highlighting two key challenges: third-party code customization and the diversity of build configurations. Xu et al.~\cite{xu2020automatic} analyzed Android kernel vulnerability patches, showing that security patches are typically smaller than non-security patches, and larger ones are often composed of several small individual fixes. Li and Paxson~\cite{li2017large} studied the life cycle of kernel security patches, finding that they are more localized but often delayed. Mu et al.~\cite{mu2018understanding} assessed the reproducibility of crowd-reported vulnerabilities, and later examined factors contributing to duplicate kernel bug reports, offering guidance for deduplication.
In contrast, our work focuses not on genuine kernel bugs or patches, but on false-positive bug reports where issues are mistakenly reported as kernel defects. We perform a detailed analysis of their root causes and affected components, and further propose an LLM-based method to automatically identify false positives to assist developers in bug triage.

\smallskip
\noindent
\textbf{Bug Report Classification Studies}.
Prior work has also explored bug report classification, aiming to distinguish bug reports from other issue types to streamline triage. Du et al.~\cite{du2021empirical} classified reports into ``Bug'' and ``Non-bug''. Herzig et al.~\cite{herzig2013s} showed that many issues labeled as bugs in repositories are actually enhancements or other categories. Building on such observations, automated classification approaches have been proposed. Zhou et al.~\cite{zhou2016combining} combined textual and structural features to separate bug reports from feature requests. Pandey et al.~\cite{pandey2017automated} applied neural network models to classify GitHub issues, outperforming traditional methods. More recently, LLMs have been applied: for example, Aracena et al.~\cite{aracena2024applying} demonstrated that GPT-based models can classify and prioritize issue reports under few-shot settings.
While these works advance general issue classification, they mostly deal with broad categories such as bug reports, feature requests, and questions. However, false-positive bug reports form a distinct class where reporters mistakenly flag non-bugs as defects. Unlike feature requests or general questions, these reports require developers to clarify misconceptions rather than fix actual faults, demanding a deeper semantic understanding of system behavior and report context. Our study explicitly targets this overlooked category in the Linux kernel, filling a gap by characterizing false positives and developing automated methods to detect them.

\smallskip
\noindent
\textbf{Kernel Fuzzing}.
The Linux kernel has been a primary target of fuzzing research, explored through several directions. The most common is system call fuzzing, where tools such as Syzkaller generate and mutate system call sequences to uncover bugs~\cite{trinity,syz,kim2020hfl,schumilo2017kafl,sun2021healer,wang2021syzvegas}. To improve effectiveness, analysis-enhanced fuzzing combines fuzzing with static analysis~\cite{pailoor2018moonshine}, symbolic execution~\cite{cadar2008klee}, dynamic analysis~\cite{sun2021healer}, enabling deeper exploration of hard-to-reach paths. Furthermore, interface-specific fuzzing extends testing to DMA~\cite{song2019periscope}, USB~\cite{peng2020usbfuzz}, and driver interrupts~\cite{hetzelt2021via}, broadening coverage across kernel subsystems.
Our work complements this line of research by shifting focus from detecting bugs to classifying reported ones. 
We address this challenge by systematically evaluating LLM-based techniques for identifying false-positive reports and explaining their causes, aiming to reduce developer effort while preserving the rigor of kernel debugging.

 \section{Data Preparation}
\label{sec:data}

\textbf{Studied Linux Bug Trackers.}
In this work, we focus on two representative platforms for Linux kernel bug tracking (i.e., \textbf{Bugzilla}~\cite{bugzilla} and \textbf{Syzkaller}~\cite{syz}) given their prominence and centrality in the Linux ecosystem.
Bugzilla, a long-standing general-purpose tracker, serves as the primary venue for reporting and managing kernel issues.
With structured workflows, rich metadata, and extensive historical archives, it is indispensable for analyzing human-reported false positives.
In contrast, Syzkaller, an automated fuzzer developed by Google, continuously generates crashing inputs and files reports via its integrated dashboard. 
Equipped with reproducer programs and detailed crash metadata, it surfaces tool-driven reports that complement human submissions.
Together, these platforms illustrate the heterogeneous nature of kernel bug reporting: Bugzilla reflects real-world contexts but is prone to subjective errors, while Syzkaller offers broad coverage yet produces many spurious reports.
By jointly analyzing both, our dataset captures diverse scenarios, from user-reported cases to automated fuzz testing, enhancing the representativeness and generalizability of our findings.
Note that we deliberately exclude GitHub pull requests, as the Linux kernel community does not rely on GitHub for bug tracking or maintenance.

\smallskip
\noindent
\textbf{Data Collection and Cleaning.} For Bugzilla, we focus exclusively on \textit{resolved} bug reports, as \textit{unresolved} reports often lack sufficient diagnostic context or resolution, making it difficult to determine whether they correspond to genuine bugs or false-positive bug reports. 
Specifically, we consider reports fixed through a code change in the Linux kernel (marked as \textit{resolved: code fix} on Bugzilla) as potential genuine bugs, while reports explicitly marked as invalid by developers (marked as \textit{resolved: invalid} on Bugzilla) are treated as potential false-positive bug reports. 
In total, we collected a total of 1,906 Bugzilla reports, including 1,478 potential genuine bugs and 428 potential false positives.
For Syzkaller, we adopt a similar strategy. Reports labeled as \textit{fix bisection done} are treated as potential genuine bugs, since this status indicates that a concrete patch resolving the issue was identified and confirmed by the kernel developers. Conversely, reports marked as \textit{invalid} are considered potential false positives. However, we observed that many invalid reports lacked developer responses beyond the initial submission, providing insufficient information to reliably determine whether they were indeed false positives. To ensure data quality, we filtered out such reports without any developer discussion. In total, we collected 1,766 Syzkaller reports, including 437 potential genuine bugs and 1,329 potential false positives.
Note that we limited our data collection to a specific time window, from January 2020 to July 2025 (over five years), because manually analyzing each bug is labor-intensive, making it impractical to examine the entire historical dataset.

\smallskip
\noindent
\textbf{Annotation of False-Positive Bug Reports.}
In the previous step, we distinguished potential genuine bugs from false-positive reports using the system-provided status labels. However, these labels are not entirely reliable. For example, Bugzilla’s status field can sometimes be modified by the reporter, which introduces noise into the dataset. To address this, we applied an additional annotation phase to ensure data quality.

Specifically, we adopted a two-stage labeling process that combines automatic pre-annotation with large language models (LLMs) and manual verification.
\uline{(1) LLM-based pre-annotation.} In this stage, we drew inspiration from prior work on voting strategies~\cite{shen2021comprehensive} and LLM-as-a-judge methods~\cite{gu2024survey}. We employed two advanced LLMs, DeepSeek~\cite{deepseek} and Qwen~\cite{qwen}, to independently provide preliminary labels along with justifications for their decisions, which served as references for subsequent human inspection. 
Reports consistently labeled as false positives by both models were treated as false-positive candidates, while those consistently labeled as genuine bugs were marked as genuine bug candidates. 
By offering candidate labels and rationales for clear-cut cases, this automated pre-annotation substantially reduced the manual workload.
\uline{(2) Manual verification.} To guarantee the reliability of the dataset, all auto-labeled entries (both genuine bugs and false positives) underwent a structured manual validation process. Two authors independently examined each report, following a systematic annotation protocol inspired by prior work~\cite{wang2023conversations}. The evaluation considered the issue content, developer responses, error traces, and resolution outcomes. Reports were confirmed as genuine bugs if they described kernel defects acknowledged and fixed by developers. In contrast, they were labeled as false positives if the issues arose from user misuse, invalid inputs, misconfigurations, or misunderstandings of kernel behavior, without any developer-side code changes. Reports lacking sufficient discussion, diagnostic detail, or conclusive resolution were excluded from the final dataset.
To assess labeling consistency, we measured inter-rater agreement using Cohen’s Kappa coefficient~\cite{cohen1960coefficient, cohen2}. In an initial calibration round covering 3\% of the data (120 reports), we observed substantial agreement (Cohen’s Kappa = 0.71), which led to a discussion to refine labeling guidelines and resolve ambiguities. After adjustments, inter-rater agreement improved significantly, reaching 0.94 in the next 15\% of the dataset (600 reports). The final round achieved strong agreement (Cohen’s Kappa > 0.95), indicating high labeling consistency.

\begin{table}[t]
\centering
\renewcommand{\arraystretch}{1.2}
\caption{Statistical Information on Datasets}
\label{tab:dataset}
\resizebox{0.75\linewidth}{!}
{
\begin{threeparttable}

\begin{tabular}{l|c|cccc}
\toprule

\textbf{Source}    & \textbf{Duration} &     \textbf{Genuine Bug} & \textbf{False-Positive} & \textbf{Total} & \textbf{False-Positive Ratio} \\
\midrule
\textbf{Bugzilla}  & \multirow{2}{*}{2020-01 $\sim$ 2025-07}  & 1,072                 & 303                     & 1,375           & 22.04\%              \\
\textbf{Syzkaller} &   & 437                  & 194                     & 631            & 30.74\%               \\
\midrule
\textbf{Total}     &  2020-01 $\sim$ 2025-07  & 1,509                 & 497                     & 2,006           & 24.78\%         \\

\bottomrule
\end{tabular}
\end{threeparttable}
}
\end{table}

Through this rigorous two-stage process, we obtained a high-quality dataset for subsequent analysis. The final dataset spans a five-year period (January 2020 $\sim$ July 2025) and includes both genuine and false-positive bug reports. As summarized in Table~\ref{tab:dataset}, we collected 497 false-positive reports and 1,509 genuine bug reports, with an overall false-positive rate of 24.78\% (22.04\% in Bugzilla and 30.74\% in Syzkaller). These proportions underscore the substantial prevalence of false-positive bug reports in Linux kernel bug trackers.
Furthermore, our analysis shows a steady increase in false positives from 2021 to 2024, signaling a growing challenge for Linux kernel maintenance.

\section{Findings}

\subsection{RQ1: Developer Effort Overhead of False-Positive Bug Reports}
\label{sec:rq1}

\textbf{Approach}.
To address RQ1, we perform a quantitative comparison of developer effort involved in handling false-positive versus genuine bug reports.
Following prior studies on issue report analysis~\cite{kuramoto2024understanding,zhang2012empirical}, we extracted metadata from both Bugzilla and Syzkaller datasets, including the timestamp of report creation (\textit{open\_time}), timestamp of the last reply (\textit{last\_comment\_time}), the total number of comments, and the number of unique participants involved in each discussion thread. These metadata enable us to capture the developer effort invested in handling different types of reports.
Specifically, we define three metrics:

\begin{itemize}
\item \textbf{Number of Participants:} represents the number of unique participants engaged in the report, reflecting the breadth of developer involvement.
 \item \textbf{Number of Comments:} represents the total count of comments in a report, reflecting the level of back-and-forth interaction required for resolution.
 \item \textbf{Time to Close:} represents the duration (hours) between \textit{open\_time} and \textit{last\_comment\_time} of a report. Specifically, for Syzkaller, since a single report often contains multiple discussion threads (e.g., related to different maintainers), we compute this metric as the sum of the durations of all associated discussions. 
\end{itemize}

To determine whether developer effort differs significantly between false-positive and genuine bug reports across these metrics, we applied the Mann-Whitney U test~\cite{mcknight2010mann}, a non-parametric statistical test used to assess whether two independent samples originate from the same distribution. This test is particularly suitable for comparing ordinal data or variables that do not follow a normal distribution. We report p-values to evaluate the presence of statistically significant differences, with p < 0.05 indicating a statistically significant result. Moreover, we measured the effect size using Cliff’s Delta~\cite{cliff1}, a non-parametric metric that quantifies the degree of overlap between two distributions. Effect size is analyzed as follows: (1) $|\delta| \textless 0.147$ as Negligible, (2) $0.147 \leq |\delta | \textless 0.33$ as Small, (3) $0.33 \leq | \delta | \textless 0.474$ as Medium, or (4) $0.474 \leq |\delta |$ as Large~\cite{cliff2}. These metrics together allow us to assess not only whether the differences are statistically significant but also whether they are practically meaningful.

\begin{figure}
    \centering
    \includegraphics[width=0.95\linewidth]{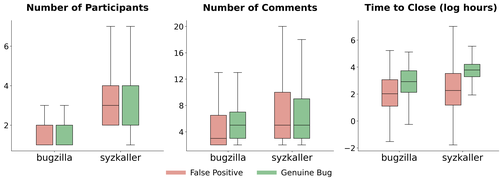}
    \caption{Comparison of False-Positive and Genuine Bug Reports in the Linux Kernel}
    \label{fig:rq1_result}
    \vspace{-10pt}
\end{figure}

\smallskip
\noindent
\textbf{Results and Analysis}.
Figure~\ref{fig:rq1_result} presents a comparative boxplot of false-positive and genuine bug reports across the three developer-effort metrics for Bugzilla and Syzkaller. The corresponding statistical test results are summarized in Table~\ref{tab:rq1_2}.

\uline{\textbf{Finding 1:} False-positive bug reports require substantial developer effort, exhibiting comparable levels of discussion and participation to genuine bugs, while still demanding non-trivial closure times. This demonstrates that triaging false positives is not a trivial task but a significant burden for maintainers.}
For the \textit{Number of Participants}, Syzkaller shows a statistically significant difference between false-positive and genuine bug reports ($p < 0.05$), though the effect size remains small ($\delta = 0.2082$). By contrast, Bugzilla exhibits no statistically significant difference ($p = 1.00, \delta=0.0001$). On average, Bugzilla false-positive reports involve 2.03 participants, compared to 1.98 participants for genuine reports, and Syzkaller false positives engage 3.59 participants versus 3.29 participants for genuine reports. In terms of medians, Bugzilla shows parity (2 vs. 2 participants), whereas Syzkaller false positives slightly exceed genuine reports (3 vs. 2 participants). This suggests that developer engagement in false-positive bug reports is at least comparable (and in some cases slightly higher) than that for genuine bugs.

For the \textit{Number of Comments}, both Bugzilla and Syzkaller exhibit statistically significant differences, though with small or negligible effect sizes. On Bugzilla, false positives average 5.85 comments compared to 5.49 for genuine bugs. On Syzkaller, false positives average only 10.70 comments, whereas genuine bugs attract 26.56 comments. 
This discrepancy largely reflects Syzkaller’s reporting workflow: genuine bugs often trigger follow-up reports, multiple bisection attempts, and extended discussions to validate reproductions, while many false positives are identified as non-defects and require fewer iterations. 
Nevertheless, the median number of comments remains close across both categories: 3 versus 5 on Bugzilla, and 5 versus 5 on Syzkaller, indicating that typically the discussion load is comparable for both false positives and genuine bugs.
\tian{While the quantitative results show that false-positive reports attract levels of participation and discussion comparable to genuine bugs, this effort is often qualitatively different in nature. Based on a manual inspection of discussion threads, we observe that maintainers frequently spend substantial time examining low-level diagnostics (e.g., KASAN reports) and repeatedly requesting missing information such as kernel configurations or reproducers. This suggests that the comparable discussion volume in false positives does not reflect productive debugging, but rather overhead induced by incomplete or ambiguous reports.}

\begin{table*}[t]
\caption{Statistical Test Results Comparing False-Positive and Genuine Bug Reports}
\centering
\small
\label{tab:rq1_2}
\renewcommand{\arraystretch}{1.2}
\resizebox{0.9\linewidth}{!}
{
\begin{threeparttable}
\begin{tabular}{l|l|ccc}
\toprule
\textbf{Metric}                         & \textbf{Project} & \multicolumn{1}{c}{\textbf{Mann-Whitney U p-value}} & \multicolumn{1}{c}{\textbf{Cliff’s Delta}} & \multicolumn{1}{c}{\textbf{Magnitude}} \\ \midrule
\multirow{2}{*}{Number of Participants} & Bugzilla         & 0.9977646                          & 0.0001                                       & Negligible                             \\
                                        & Syzkaller        & 1.216E-05                         & 0.2082                                       & Small                             \\ \midrule
\multirow{2}{*}{Number of Comments}     & Bugzilla         & 6.488E-05                      & -0.1487                                      & Small                                  \\
                                        & Syzkaller        & 0.0233996                       & -0.1126                                      & Negligible                                  \\ \midrule
\multirow{2}{*}{Time to Close (hours)}   & Bugzilla         & 5.137E-19                                            & -0.3348                                      & Medium                                 \\
                                        & Syzkaller        & 1.785E-29                                            & -0.5648                                      & Large                                  \\

\bottomrule
\end{tabular}
\end{threeparttable}
}
\end{table*}

For the \textit{Time to Close}, both Bugzilla and Syzkaller exhibit statistically significant differences with medium-to-large effect sizes ($p < 0.05$, $|\delta| > 0.33$ for Bugzilla; $|\delta| > 0.47$ for Syzkaller). As expected, false positives are generally closed faster than genuine bugs, since they do not require extensive debugging and the verification of patches. Yet, this efficiency comes at a cost: resolving false positives often demands the involvement of more developers and non-trivial discussions before reaching consensus. Furthermore, the total time investment remains substantial: false positives still account for an average of 5,448.57 hours on Bugzilla and 2,690.94 hours on Syzkaller, with median values of 103.73 and 159.57 hours, respectively.
Therefore, even though they reach closure sooner, false positives still consume significant developer effort.

\begin{tcolorbox}[colback=gray!5]
\textbf{RQ1 Summary}: 
False-positive bug reports impose a substantial overhead on kernel maintainers, comparable to that of genuine bugs. They demand considerable resolution time, involve multiple developers, and often trigger extensive discussions.
These findings support our core motivation: reducing the effort spent on identifying and closing false-positive bug reports can lead to more efficient resource allocation in Linux kernel maintenance.
\end{tcolorbox}

\subsection{RQ2: Components of False-Positive Bug Reports}
\label{sec:rq2}

\textbf{Approach.}
To locate where false-positive bug reports are most likely to occur, we use the component labels provided by Bugzilla and Syzkaller to categorize bug reports. 
Although the two systems adopt different naming conventions, their schemes are broadly aligned at the subsystem level (e.g., networking, file system, drivers). We standardized the taxonomy by taking Bugzilla’s component classification as the reference~\cite{bugzilla_cgi} and mapping Syzkaller reports into this scheme. 
\tian{This mapping was initially performed independently by two authors, drawing on component labels, Linux kernel documentation, and bug report descriptions.
Disagreements were then resolved through round-table discussions and a card-sorting process~\cite{card_sorting}, ensuring consistency across datasets and minimizing subjective bias.
Only 6.78\% of cases required discussion, indicating that most mappings were straightforward. The independent annotations achieved a Cohen’s Kappa of 0.92, reflecting strong agreement.}
Our analysis shows that the top seven components (i.e., File System, Drivers, Networking, Kernel Core, Tools, Security, and IO) account for nearly 80\% of all false positives. 
The definitions of these components are provided in Section~\ref{sec:background}.
In contrast, the remaining components (e.g., Virtualization, Timers, etc) each contribute less than 4\%.
To ensure representativeness, we focus our analysis on the top seven components in the following discussion.

\begin{figure}[t]
  \centering
  \includegraphics[width=0.9\linewidth]{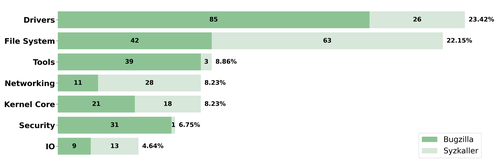}
  \caption{False-Positive Bug Reports Distribution by Component}
  \label{fig:component}
\end{figure}

\smallskip
\noindent
\textbf{Results and Analysis.}
Figure~\ref{fig:component} shows the distribution of false-positive reports across Linux kernel components, highlighting which components are most vulnerable to false-positive reports.

\uline{
\textbf{Finding 2:} File Systems and Drivers are the primary sources of false positives, underscoring a need for better validation of edge-case behaviors and hardware interactions.
}
\tian{The following explanations are hypotheses derived from our empirical observations, rather than confirmed causal conclusions, and will be further validated through developer-side studies in future work.}
As shown in Figure~\ref{fig:component}, false positives are heavily concentrated in the File System and Drivers components, which together account for more than 40\% of all cases. Within the File System, false positives may arise from complex but valid operational sequences such as cross-subsystem interactions (e.g., VFS, ext4, btrfs). These behaviors, though correct, may appear anomalous to automated tools or less experienced developers. In Drivers, false positives \tian{may frequently result from} interactions with diverse hardware. Incomplete specifications or tool misjudgment can cause normal device-specific behavior to be mistaken for errors. These observations point to practical improvements: for File Systems, better modeling of valid operation sequences and usage-aware test oracles could help reduce false positives; for Drivers, incorporating hardware-aware validation and improving specification coverage may reduce false positives.

\uline{
\textbf{Finding 3:} Human-reported false positives in Bugzilla are more prevalent in Drivers, Tools, and Security compared to Syzkaller, reflecting human misunderstandings that could be mitigated through clearer guidance. 
}
Bugzilla shows a higher rate of false positives in Drivers, Tools, and Security compared to Syzkaller (155 vs. 30), while both sources produce similar numbers in Kernel Core (21 vs. 18) and IO (9 vs. 13). This distribution aligns with Bugzilla’s user-driven nature: users frequently interact with Tools and Security features and are more likely to report perceived anomalies in these areas. However, without deep kernel knowledge, they may misinterpret normal error handling, security policies, or tool limitations as bugs.
By contrast, fuzzers like Syzkaller seldom exercise these userspace-facing components extensively.
These results suggest that false positives in Bugzilla often stem from human error during frequent interaction. Improved documentation, clearer diagnostic messages, and better triage support could help reduce such false positives. 

\uline{
\textbf{Finding 4:} Syzkaller produces more false positives in Networking and File Systems, highlighting limitations in automated testing design.
}
Syzkaller contributes more false positives in Networking and File System, where it can systematically test but humans rarely touch at scale. Its fuzzing approach uncovers many rare but valid behaviors, such as unusual socket state transitions or obscure filesystem mount/unmount sequences, that are flagged as bugs. By contrast, Bugzilla does not show the same concentration because human reporters rarely encounter these corner cases. 
These results highlight that Syzkaller’s false positives come from weaknesses in automated testing rather than misunderstanding, and could be mitigated by improving fuzzing pipelines with stronger semantic checks, impact-based filtering, or developer feedback loops.

\begin{tcolorbox}[colback=gray!5]
\textbf{RQ2 Summary}: 
False-positive bug reports are concentrated in certain components of the Linux kernel. File System and Drivers components account for the largest proportions, which together account for over 40\% of all cases. 
\end{tcolorbox}

\subsection{RQ3: Root Causes of False-Positive Bug Reports}
\label{sec:rq3}

\textbf{Approach.}
To identify the root causes of false-positive bug reports, we manually analyzed a dataset of \fpnum{} reports (194 from Syzkaller and 303 from Bugzilla) as described in Section~\ref{sec:data}. 
We began by referencing taxonomies from prior studies on false-positive bug reports in traditional software~\cite{laiq2022early, laiq2023data}, using them as an initial taxonomy. Through an iterative annotation process, we progressively refined this taxonomy to capture Linux kernel–specific patterns following an open-coding approach~\cite{corbin1990grounded}. 
This allowed us to better reflect real-world Linux kernel false-positive bug reports.
For each entry, two authors independently analyzed descriptions and developer discussions to determine root causes. 
To ensure labeling consistency, we measured inter-rater agreement using Cohen’s Kappa coefficient. An initial round on 10\% of the data (50 false-positive bug reports) yielded moderate agreement (Cohen’s Kappa = 0.51), prompting a calibration discussion. Agreement improved to 0.87 in the next 10\%, and subsequent rounds exceeded 0.95 after resolving inconsistencies. Remaining disagreements were adjudicated by a third author to ensure accurate labeling of all false-positive bug reports. 

Drawing from an in-depth analysis of false-positive bug reports of Linux kernel, we devise a taxonomy of root causes behind them, as outlined below:

\textbf{(1) External Dependency Issues.} Since kernel execution often depends on external components, problems in peripheral domains are frequently misclassified as kernel bugs. We further define three subtypes: Hardware Issues (e.g., malfunctioning devices), Userspace Dependency Issues (e.g., defects in userspace applications and libraries), and Outdated Firmware or Firmware Issues (e.g., device firmware incompatibilities). These dependencies blur the boundary of responsibility between kernel and external modules, complicating root cause attribution.

\textbf{(2) Misunderstanding of Features or Limitations.} This category refers to misinterpretations of expected kernel behavior, where correct but non-intuitive behaviors are reported as bugs. We further define three subtypes: Implicit Behaviour Misunderstanding (e.g., misinterpreting concurrency, scheduling, or timing effects that are by design), Limitation Unawareness (e.g., failure to recognize intentional restrictions, such as resource caps), and Semantic Misunderstanding (e.g., incorrect interpretations of API specifications). These cases highlight how subtle gaps between user understanding and intended kernel behavior can lead to false bug reports.

\textbf{(3) Incorrect Environment Configuration.} This category refers to failures caused by misconfigured or improper system environments, where the kernel is blamed for issues that actually stem from setup mistakes. We further define two subtypes: Software Environment Configuration (e.g., inconsistent kernel builds, misconfigured system parameters) and Hardware Environment Configuration (e.g., incompatible CPU architectures, missing device drivers). 
Such errors often emerge due to the inherent complexity of configuring and integrating multiple kernel components.

\textbf{(4) Incorrect Usage.} This category refers to failures when the kernel is used in ways that deviate from intended practices. We further define two subtypes: Unsupported Invocation (e.g., enabling legacy options or using obsolete APIs) and Incorrect Operation (e.g., invoking privileged commands without required permissions). These reports typically reflect reporter unfamiliarity with appropriate usage boundaries.

During the manual classification of root causes, we further annotated the observable sub-root causes associated with each category. To systematically identify and organize these subcategories, we employed a card sorting method~\cite{card_sorting}, enabling us to uncover common root causes and improve our understanding of how different root causes manifest in false-positive bug reports.
\tian{Specifically, for ambiguous cases, we used the developers’ resolution in the discussion as the authoritative signal for determining root causes. Specifically, if an issue was resolved through hardware-related actions (e.g., disabling faulty components), it was classified as a Hardware Issue; if it was resolved by modifying build-time or runtime settings, it was classified as a Software Configuration Issue.
For example, in \textit{Bug\#206141}\footnote{\url{https://bugzilla.kernel.org/show_bug.cgi?id=206141}}, the fix appears at first glance to be a configuration change, but it actually disables specific GPU rings and was therefore attributed to defective hardware behavior. While a small number of reports could plausibly fit multiple categories, such cases account for only 4.8\% of the dataset, indicating that category overlap is limited.
}

\begin{figure}[t]
  \centering
  \includegraphics[width=0.95\linewidth]{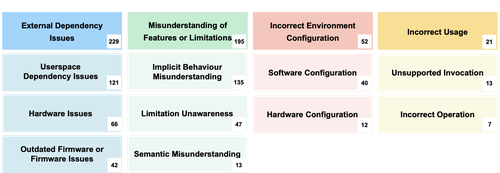}
  \caption{False-Positive Bug Reports Root Causes}
  \label{fig:cause}
\end{figure}

\smallskip
\noindent
\textbf{Results and Analysis}.
Figure~\ref{fig:cause} presents the distribution of identified root causes and their associated subcategories for false-positive bug reports in the Linux kernel. Each root cause highlights distinct challenges related to kernel usability and ecosystem design. 
To complement this overview, Tables~\ref{tab:rootcause} provides detailed statistical summaries of each subcategory.

\uline{
\textbf{Finding 5:} External dependency issues are the most frequent root cause of false-positive bug reports (46.1\%), with more than half (52.8\%) involving userspace dependency issues.
}
From Figure~\ref{fig:cause}, external dependency issues emerge as the dominant source of false positives, accounting for 229 out of the 497 cases in our dataset (46.1\%). This prevalence reflects the inherent role of the Linux kernel as an intermediary between userspace applications and diverse hardware devices, which naturally entangles it with external components such as device firmware, various hardware, and userspace libraries. Consequently, when failures occur at these boundaries, they are frequently misattributed to kernel defects even though the underlying root cause lies outside the kernel. The kernel’s rapid evolution and frequent release cycles further exacerbate this issue, as new features or changes may surface latent incompatibilities in external dependencies.
From Table~\ref{tab:rootcause}, we observe that external dependency issues are markedly more common in Bugzilla than in Syzkaller. This contrast arises because Bugzilla reports come from end users operating in highly diverse and uncontrolled environments, where hardware variability, userspace configurations, and firmware mismatches are commonplace. 
In contrast, Syzkaller runs in controlled fuzzing setups with standardized and virtualized environments that minimize such external factors. As a result, Syzkaller false positives are less likely to stem from dependency-related problems.

To further understand these cases, we categorize external dependency–related false positives into three subcategories: \textit{Userspace Dependency Issues} (121 cases, e.g., mismatches between kernel APIs and glibc or other system libraries), \textit{Hardware Issues} (66 cases, e.g., device-specific quirks leading to spurious kernel error messages), and \textit{Firmware Issues} (42 cases, e.g., outdated or buggy firmware incorrectly triggering kernel warnings). Among these, Userspace Dependencies Issues dominate, accounting for over half of the external dependency issues (52.8\%), highlighting how tightly the kernel depends on the correctness and compatibility of userspace components.

\begin{figure}[t]   
  \centering  
    \subfloat[Userspace Dependency Issue]
  {
      
      \label{fig:case2}\includegraphics[width=0.45\textwidth]{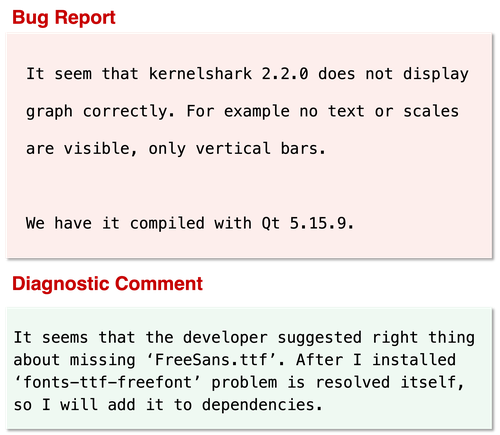}
      
  }
  \subfloat[Implicit Behavior Misunderstanding] 
  {
      \label{fig:case1}\includegraphics[width=0.45\textwidth]{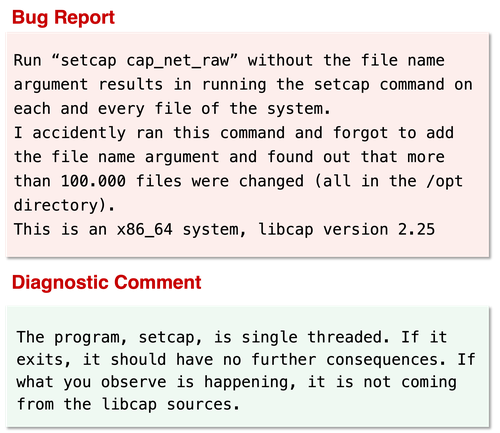}
  }
  \caption{Illustrative Examples}   
  \label{fig:illustrative_example}           
  \vspace{-10pt}
\end{figure}

\textbf{Illustrative Example.}
Figure~\ref{fig:case2} illustrates a false-positive bug report in Bugzilla caused by Userspace Dependencies Issues. The reporter observed that KernelShark 2.2.0 failed to render graphs correctly, showing only vertical bars without text or scales, and initially suspected a kernel defect. The issue was reported with accompanying screenshots and trace data, compiled in an environment using Qt 5.15.9. Upon developer investigation, however, the visualization problem was traced to the Qt libraries. 
Specifically, the bug disappeared once Qt was rebuilt from source.
This case highlights how mismatches or inconsistencies in userspace dependencies can lead reporters to misattribute failures to the kernel. 

\begin{table}[t]
\centering
\caption{Distribution of Root Causes and Subcategories}
\label{tab:rootcause}
\resizebox{0.95\linewidth}{!}
{
\begin{threeparttable}
\begin{tabular}{l|l|ccc}
\toprule
\textbf{Category} & \textbf{Subcategory} & \textbf{Bugzilla} & \textbf{Syzkaller} & \textbf{Total} \\
\midrule
\multirow{4}{*}{External Dependency Issues} 
  & Hardware Issues                     & 45 & 21 & 66 \\
  & Userspace Dependency Issues         & 85 & 36 & 121 \\
  & Outdated Firmware or Firmware Issues& 41 & 1  & 42 \\
  & Total                               & \textbf{171} & \textbf{58} & \textbf{229} \\
\midrule
\multirow{4}{*}{Misunderstanding of Features or Limitations} 
  & Limitation Unawareness              & 22 & 25 & 47 \\
  & Semantic Misunderstanding           & 6  & 7  & 13 \\
  & Implicit Behavior Misunderstanding  & 53 & 82 & 135 \\
  & Total                               & \textbf{81} & \textbf{114} & \textbf{195} \\
\midrule
  \multirow{3}{*}{Incorrect Environment Configuration} 
  & Hardware Environment Configuration  & 10 & 2 & 12 \\
  & Software Environment Configuration  & 31 & 9 & 40 \\
  & Total                               & \textbf{41} & \textbf{11} & \textbf{52} \\
\midrule
\multirow{3}{*}{Incorrect Usage} 
  & Unsupported Usage                   & 6 & 8 & 13 \\
  & Incorrect Operation                 & 4 & 3 & 7 \\
  & Total                               & \textbf{10} & \textbf{11} & \textbf{21} \\

\bottomrule
\end{tabular}
\end{threeparttable}
}
 \vspace{-10pt}
\end{table}

\uline{
\textbf{Finding 6:} Misunderstanding of Features or Limitations accounts for 39.2\% of false-positive bug reports, primarily driven by implicit behavior misunderstandings.
}
From Figure~\ref{fig:cause}, 195 out of the 497 false-positive cases stem from misinterpretations of kernel functionality, making this the second most frequent cause. These errors often arise when reporters, unfamiliar with the kernel’s internal semantics or design assumptions, misjudge expected behaviors as defects. For example, many kernel components deliberately adopt conservative policies (e.g., strict error checking or defensive warnings) that may appear to users as signs of bugs. Similarly, backward-incompatible changes or long-standing limitations may be mistaken for regressions, especially when users lack awareness of related documentation or commit messages. 
As a result, misunderstanding kernel behavior frequently leads to spurious reports that divert developer effort away from genuine issues.

We further divide this category into three subcategories: \textit{Limitation Unawareness} (47 cases, e.g., failure to recognize that certain system calls are intentionally restricted under specific contexts), \textit{Semantic Misunderstanding} (13 cases, e.g., misinterpreting error codes or log messages as kernel crashes), and \textit{Implicit Behavior Misunderstanding} (135 cases, e.g., treating implicit behavior such as performance optimizations or lazy initializations as bugs). Among these, implicit behavior misunderstanding dominates, representing 69.2\% of cases in this category. This suggests that the kernel’s complex and often implicit runtime behaviors, which are rarely visible without in-depth knowledge, remain a major source of confusion for users.

\textbf{Illustrative Example} Figure~\ref{fig:case1} shows a false-positive bug report caused by Implicit Behaviour Misunderstanding. A user mistakenly believed that running setcap cap\_net\_raw without specifying a target file would trigger a system-wide capability change, affecting all files. The user described this as a severe defect, reporting that more than 100,000 files was modified. However, developers clarified that the behavior was misinterpreted: the setcap tool is single-threaded and does not recursively alter arbitrary files without explicit instruction. The observed outcome was likely caused by external factors, not by the libcap sources themselves.
This case exemplifies how implicit assumptions about system commands such as expecting a safeguard or warning prompt when required arguments are missing can lead users to misinterpret intended behavior as a bug.

\begin{table}[t]
\centering
\caption{Stage-wise Root Cause Proportions}
\label{tab:rq3_rc}
\resizebox{0.9\linewidth}{!}
{
\begin{threeparttable}
\begin{tabular}{l|
>{\centering\arraybackslash}p{3.0cm}
>{\centering\arraybackslash}p{3.5cm}
>{\centering\arraybackslash}p{3.8cm}
>{\centering\arraybackslash}p{2.8cm}}
\toprule
            & External Dependency Issues & Misunderstanding of Features or Limitations & Incorrect Environment Configuration & Incorrect Usage \\
\midrule
Drivers     & 47.75\%                                        & 29.73\%              & \textbf{19.82\%}                                & 2.70\%              \\
File System & 24.76\%                               & \textbf{61.90\%}                       & 7.62\%                                 & 5.71\%             \\
Kernel Core & 48.72\%                                        & 38.46\%              & 7.69\%                                 & 5.13\%             \\
Networking  & 46.15\%                               & 51.28\%                       & 0                                      & 2.56\%             \\
Security    & \textbf{75.00\%}                                         & 12.50\%               & 6.25\%                                 & 6.25\%             \\
Tools       & 45.24\%                                        & 35.71\%              & 11.90\%                                 & \textbf{7.14\%}          \\
IO  & 63.64\%                        & 36.36\%                                          & 0                                    & 0               \\
\bottomrule
\end{tabular}
\end{threeparttable}
}
\vspace{-10pt}
\end{table}

\uline{
\textbf{Finding 7:}False-positive bug reports show component-specific root causes, suggesting a \linebreak relationship between root causes and kernel components.
}
Since different kernel components provide distinct functionalities and interact with diverse subsystems, certain types of false positives are more likely to emerge in specific components. 
To investigate this, we further analyze the relationship between root causes and kernel components, and employ Pearson’s Chi-Square test~\cite{plackett1983karl} to determine whether there is a statistically significant difference. 
Table~\ref{tab:rq3_rc} then presents the proportion distribution of false-positive root causes across major kernel components, enabling us to identify component-specific patterns.

First, \textit{External Dependency Issues} dominate in components with strong reliance on userspace libraries, firmware, or hardware devices. For example, the Security component shows the highest concentration (75\%), as security features often depend on cryptographic libraries, hardware support, or userspace configuration tools. Similarly, IO (63.64\%) subsystems show high dependency-related false positives, as IO relies on heterogeneous storage hardware and drivers.

Second, \textit{Misunderstanding of Features or Limitations} is particularly prominent in the File System (61.9\%) and Networking (51.28\%) components. File systems often expose subtle semantics (e.g., delayed writes or concurrency behaviors) that can be counter-intuitive to users, leading to frequent misclassification of expected behaviors as bugs. Similarly, networking involves complex protocol interactions and timing dependencies, which are easily misinterpreted as erroneous behavior.

Third, \textit{Incorrect Environment Configurations} are more pronounced in Drivers (19.82\%) and Tools (11.9\%). These components are sensitive to build options, kernel configuration flags, or environment mismatches. For example, driver-related reports often result from hardware-specific setups or missing firmware, while toolchain-related reports can arise from inconsistent software configurations.

Finally, \textit{Incorrect Usage} contributes only a small fraction across all components, rarely exceeding 7\%. This suggests that while user misuse or unsupported operations do occur, they are not the predominant source of false positives compared to misunderstandings and dependency-related issues.
Consistent with these observations, Pearson’s Chi-Square test (\textit{p-value} = $2.57 \times 10^{-6}$) confirms a significant correlation between root causes and Linux components.

Overall, these findings demonstrate that the root causes of false positives are not uniformly distributed across kernel components. Instead, they closely follow the component’s degree of dependency on external components and the subtlety of its semantics. This insight points toward the need for component-aware mitigation strategies. 

\begin{tcolorbox}[colback=gray!5]
\textbf{RQ3 Summary:}
The majority of false-positive bug reports in the Linux kernel arise from \textit{External Dependency Issues} (46.1\%) and \textit{Misunderstanding of Features or Limitations} (39.2\%). Our component-level analysis further reveals that each component tends to be associated with specific dominant root causes. For example, dependency issues dominate in Security and IO, whereas misunderstandings are most prevalent in File Systems and Networking.
\end{tcolorbox}

\subsection{RQ4: Can LLMs Help Mitigate False-Positive Bug Reports?}
\label{sec:rq4}

\textbf{Approach.}
To investigate whether large language models (LLMs) can automatically distinguish false-positive bug reports from genuine bugs, we conducted a systematic evaluation of LLM-based classification strategies. In particular, we examined four commonly used prompting strategies: zero-shot~\cite{zero_shot}, few-shot~\cite{few_shot}, chain-of-thought (CoT)~\cite{cot}, and retrieval-augmented generation (RAG)~\cite{rag}. For each bug report, the LLM was prompted to output a binary classification label (\textit{false positive} or \textit{genuine bug}) and a brief explanation. 
The prompt design was guided by our prior insights into the common root causes of false-positive bug reports and their typical manifestations across kernel components. 
Based on these insights, we embedded targeted guidelines into the prompts to provide domain-specific context.
Below, we describe each strategy in detail:
\begin{itemize}[leftmargin=10pt]
    \item \textbf{Zero-shot.} \uline{(1) Basic zero-shot:} The model was given only the definitions of a genuine bug and a false-positive bug report, the raw report (title and description), and an instruction to classify it. Note that the targeted guidelines were not included, aiming to assess the model’s out-of-the-box capability for distinguishing false-positive bug reports from genuine bugs. \uline{(2) Enhanced zero-shot:} This variant extends the basic setting by adding the guidelines derived from our earlier findings, summarizing common component distributions and root causes of false positives to provide the model with additional domain context.
    
    \item \textbf{Few-shot.} To provide contextual guidance, we further augmented the prompt with two representative examples of annotated bug reports (balanced between genuine and false positives). These examples were carefully selected from our manually labeled dataset and included both the classification label and the rationale. By illustrating the decision-making process and expected output format, this strategy enables the model to adapt more effectively to the classification task.
    
    \item \textbf{CoT.} We asked the model to produce explicit reasoning steps before giving a final label. The prompt guided the model through four focused checks: \uline{(1) Component analysis:} identify the kernel subsystem implicated by the report and note any component-specific patterns (e.g., driver reports historically show more false positives on Bugzilla). \uline{(2) Dependency check:} look for references to userspace libraries, external packages, firmware, or environment issues that could indicate an external root cause. \uline{(3) Feature interpretation:} assess whether the described behavior matches intended or documented kernel behavior, or instead reflects a misunderstanding of implicit behavior. \uline{(4) Integrative decision:} combine the evidence from the previous steps to decide whether the report is likely a genuine bug or a false positive, then output a short explanation and the final label.  
    This step-by-step prompt is intended to make the model’s reasoning more explicit and interpretable, guiding it to weigh multiple cues rather than rely on a single token or phrase.

    \item \textbf{RAG.} We enhanced the prompt with similar previously labeled bug reports retrieved from our dataset via semantic similarity search (using semantic embeddings). 
    To maintain balance, we retrieved one genuine bug and one false positive per query. By exposing the model to comparable historical cases, RAG allows it to reason by analogy, potentially improving both classification accuracy and consistency.
\end{itemize}

We employed DeepSeek-V3~\cite{deepseek} (2024-12-26 version) as the underlying classification LLM, due to its strong performance on a wide range of natural language understanding, code reasoning, and classification tasks~\cite{liu2024deepseek,telpa,teut, clast, ase_study, wiseut}. To ensure reproducibility and determinism, we fixed the temperature parameter to 0. All other hyperparameters (e.g., top-p, maximum tokens, stop sequences) were left at their default API settings. These configurations were applied uniformly across all prompting strategies to ensure fair comparison. 
For semantic retrieval in the RAG setting, we used Qwen-embedding-0.6B, which has demonstrated strong performance on embedding-based semantic search~\cite{qwen_embedding}.
To build the knowledge base, we leveraged all collected Bugzilla reports (1,375 reports in total, including 1,072 genuine bugs and 303 false positives). For each report to be classified, we retrieved the most similar cases (except itself) from this knowledge base using cosine similarity over their embeddings.
In our experiments, we focused exclusively on Bugzilla reports, both for building the knowledge base and for evaluation. This choice balances cost and comprehensiveness while emphasizing real-world user-reported issues, which more directly reflect the challenges faced by users and developers in kernel maintenance.

\smallskip
\noindent
\textbf{Results and Analysis}.
Table~\ref{tab:rq4} summarizes the performance of the five prompting strategies (i.e., basic zero-shot, enhanced zero-shot few-shot, CoT and RAG) on the task of distinguishing false-positive bug reports from genuine bugs. Our analysis yields two main findings.

\uline{\textbf{Finding 8:} Performance of different prompting strategies varies, with retrieval-based prompting achieving the best results.}
Basic zero-shot prompting achieves only moderate performance (accuracy = 0.70, F1 = 0.81), highlighting the difficulty of interpreting highly technical reports without task-specific context. Because LLMs lack prior exposure to the terminology and practices of the Linux kernel community, they often fail to capture the subtle distinctions between genuine bugs and false positives. When augmented with targeted notices derived from our empirical findings, enhanced zero-shot prompting produces a slight improvement (accuracy = 0.71, F1 = 0.81), suggesting that domain-specific guidance helps the model recognize false positives, albeit with a small reduction in recall. Further gains are achieved with few-shot prompting, where representative labeled examples illustrate the expected reasoning and output format. This strategy yields higher effectiveness (accuracy = 0.73, F1 = 0.83), indicating that concrete examples enable the model to better align with task requirements.

Contrary to expectations, CoT prompting does not improve effectiveness and instead performs worse than other strategies (accuracy = 0.67, F1 = 0.77). Although CoT encourages explicit intermediate reasoning, kernel bug reports often include lengthy stack traces, low-level logs, and environment-dependent details that overwhelm the model. As a result, the model frequently overfits to superficial cues or hallucinates causal chains, especially when misconfigurations are subtle and not explicitly stated. In such cases, step-by-step reasoning introduces spurious justifications, leading to misclassification. This indicates that structured reasoning is not inherently beneficial in technical domains unless grounded in reliable domain-specific evidence.

Among all evaluated strategies, RAG shows the strongest performance, with an accuracy of 0.81 and an F1-score of 0.88. Most notably, it achieves the highest recall (0.91), meaning it correctly identifies nearly all genuine bug cases. This is especially critical because misclassifying a genuine bug as a false positive could result in the issue being dismissed prematurely, delaying essential fixes in the kernel. 
By retrieving and presenting semantically similar, previously labeled bug reports, RAG enriches the model’s input with contextual cues and domain-specific knowledge. 
\tian{Specifically, comparing new reports with prior genuine and false positive cases gives the model context to resolve ambiguities that the report alone cannot reveal, leading to more accurate predictions.}
As a result, retrieval effectively enhances the model’s predictions, bridging the gap between generic LLM reasoning and the specialized knowledge required for Linux kernel triage.
\tian{Across false-positive categories, the model performs comparably overall but struggles with External Dependency Issues, likely because such cases depend on external states (e.g., hardware or environment) not observable from the report, leaving the model without sufficient evidence. In contrast, Incorrect Usage cases are easier to detect, as misuse or interface violations are often explicit in the report. These findings suggest that providing structured contextual information such as hardware details or configuration parameters could further improve the accuracy of false-positive identification.}

\tian{Beyond predictive performance, RAG also offers practical benefits for maintainers: the false-positive reports it correctly identifies have an average closure time of 336.2 days, indicating substantial developer effort that could be saved through early identification. In practice, the classifier can be integrated as a lightweight pre-analysis step in existing bug-tracking systems, automatically providing an initial assessment and justification upon report submission. With a small inference overhead (1.8s per report), it can be deployed to help maintainers prioritize likely false positives.}

\begin{table}[t]
\centering
\caption{Performance of LLM-based prompting strategies for classifying false-positive bug reports.}
\label{tab:rq4}
\resizebox{0.7\linewidth}{!}
{
\begin{threeparttable}
\begin{tabular}{l|cccc}
\toprule
\textbf{Prompting Strategy} & \textbf{Accuracy} & \textbf{Precision} & \textbf{Recall} & \textbf{F1-score} \\
\midrule
\textbf{Basic zero-shot}      & 0.70 & 0.80 & 0.83 & 0.81 \\
\textbf{Enhanced zero-shot}   & 0.71 & 0.82 & 0.80 & 0.81 \\
\textbf{Few-shot}             & 0.73 & 0.83 & 0.83 & 0.83 \\
\textbf{CoT}                  & 0.67 & 0.84 & 0.72 & 0.77 \\
\textbf{RAG}                  & \textbf{0.81} & \textbf{0.85} & \textbf{0.91} & \textbf{0.88} \\
\bottomrule
\end{tabular}
\end{threeparttable}
}
\vspace{-10pt}
\end{table}

\uline{
\textbf{Finding 9:} LLMs provide generally plausible but occasionally superficial explanations.
}
Our evaluation extended beyond classification accuracy to assess the explanatory capabilities of LLMs by examining the justifications generated for correctly identified false-positive bug reports. Specifically, we conducted a manual evaluation in which the first author reviewed the explanations produced by the RAG prompting approach and assessed whether they were both plausible and factually grounded. 
\tian{To assess the reliability of this evaluation, we added a second annotator and measured inter-annotator agreement using Cohen’s Kappa. The resulting score (0.92) indicates a high level of agreement, suggesting that subjectivity in judging explanation quality is limited.}
Overall, 85.29\% of these justifications were deemed satisfactory. The overall quality of explanations suggests that LLMs can serve as interpretable assistants, providing developers with a starting point for validation while streamlining the triage of false-positive bug reports.
For example, in \textit{Bug\#219057}\footnote{\url{https://bugzilla.kernel.org/show_bug.cgi?id=219057}}, the issue was ultimately traced to faulty hardware: a malfunctioning Bluetooth dongle from the vendor Ugreen. Replacing it with a dongle from another vendor resolved the problem, confirming it was hardware-related rather than a kernel defect. 
The LLM aligned with this diagnosis, explaining that the report likely stemmed from \textit{``a specific Bluetooth dongle (Ugreen) not working properly, due to hardware compatibility or driver issues rather than a kernel bug.''}

\textbf{Data Leakage Analysis}
To evaluate the risk of data leakage, we further conducted a controlled evaluation: since DeepSeek-V3’s official training cutoff date is July 1, 2024, we restricted our dataset to bug reports submitted after this cutoff. We divided the dataset into two subsets: (1) a leak-prone subset consisting of bug reports publicly available before the cutoff, which totaling contain \textbf{1,271} entries, including \textbf{993} genuine bug and \textbf{278} false positives. (2) a no-leak subset containing \textbf{104} bug reports, including \textbf{79} genuine bugs and \textbf{25} false positives.
Table~\ref{tab:rq4_leak} summarizes the performance of different prompting strategies across the leak-prone and no-leak subsets. Overall, RAG achieves the best results on both subsets, with accuracy (0.80 vs. 0.79) and F1-score (0.87 vs. 0.87) consistently outperforming other approaches. Notably, performance trends are consistent across leak-prone and no-leak subsets, with no evidence of leakage-driven bias.
We further conducted a paired Wilcoxon signed-rank test~\cite{woolson2007wilcoxon} at the 0.05 significance level, which yielded no statistically significant difference between the two subsets (all \textit{p-values} > 0.05).

\begin{table}[t]
\centering
\caption{Performance Comparison on Leak-Prone vs. No-Leak Subsets}
\label{tab:rq4_leak}
\resizebox{0.9\linewidth}{!}
{
\begin{threeparttable}
\begin{tabular}{l|cccc|cccc}
\toprule
 \multirow{2}{*}{\textbf{Prompting Strategy}}                 & \multicolumn{4}{c|}{\textbf{Leak-Prone}}           & \multicolumn{4}{c}{\textbf{No-Leak}}              \\

                  & Accuracy & Precision & Recall & F1 score & Accuracy & Precision & Recall & F1 score \\
\midrule
\textbf{Basic zero-shot}   & 0.69     & 0.78      & 0.82   & 0.80     & 0.70     & 0.85      & 0.73   & 0.79     \\
\textbf{Enhance zero-shot} & 0.70     & 0.81      & 0.80   & 0.80     & 0.75     & 0.83      & 0.85   & 0.84     \\
\textbf{Few-shot}          & 0.72     & 0.81      & 0.83   & 0.82     & 0.75     & \textbf{0.89}      & 0.78   & 0.83     \\
\textbf{CoT}               & 0.66     & 0.82      & 0.71   & 0.76     & 0.67     & 0.83      & 0.72   & 0.77     \\
\textbf{RAG}               & \textbf{0.80} & \textbf{0.84} & \textbf{0.91} & \textbf{0.87}     & \textbf{0.79}     & 0.84      & \textbf{0.90}   & \textbf{0.87}    \\
\bottomrule
\end{tabular}
\end{threeparttable}
}
\vspace{-10pt}
\end{table}

\begin{tcolorbox}[colback=gray!5]
\textbf{RQ4 Summary:}  
LLMs demonstrate strong potential in supporting the triage of Linux kernel bug reports. Among the evaluated strategies, RAG yields the best performance, reaching 81\% accuracy, 85\% precision, and 91\% recall. Beyond classification, LLMs also provide generally reliable explanations, with 85.29\% of generated rationales judged to be accurate. These results suggest that LLMs can not only improve efficiency in distinguishing false positives from genuine bugs but also enhance transparency in automated triage.
\end{tcolorbox}

\section{Discussions}
\label{sec:discussion}

\subsection{Robustness and Generalizability}
\tian{
To contextualize the effectiveness of LLM-based methods, we also compare against lightweight baselines commonly used in bug-report classification, namely logistic regression and kNN~\cite{terdchanakul2017bug}. The results (Table~\ref{tab:cv}) show that these approaches perform substantially worse. This gap highlights the limitations of shallow, feature-based models in capturing the semantic complexity of kernel bug reports, and further motivates the use of LLMs for false-positive identification.}
\tian{We further evaluated the generalizability of our approach across different foundation models to confirm that our findings are not specific to a single architecture. By replicating the experiments with a different LLM (GPT-4o), we observed results comparable to our primary backbone, confirming the findings' robustness to LLM choice. These results demonstrate the model-agnostic nature of our framework. }
\tian{To assess robustness to data partitioning, we conduct 5-fold cross-validation (80\% retrieval / 20\% testing, no overlap), where our RAG-based approach consistently outperforms baselines and matches main results. A temporal split (pre-2023.9.1 for retrieval, post-2023.9.1 for testing) yields similar effectiveness, further demonstrating robustness to different splits and temporal shifts.}

\begin{table}[t]
\centering
\caption{Robustness and comparative analysis of classification performance.}
\label{tab:cv}
\resizebox{0.75\linewidth}{!}
{
\begin{threeparttable}

\begin{tabular}{l|l|cccc}
\toprule

                &\textbf{method/strategy}    & \textbf{accuracy} & \textbf{precision} & \textbf{recall} & \textbf{F1 score}     \\
\midrule

\textbf{\multirow{2}{*}{traditional method}} &\textbf{logistic regression} & 0.42   & 0.87    & 0.30 & 0.45 \\
                                           &\textbf{k-nearest neighbors} & 0.51   & 0.80    & 0.50  & 0.61 \\
\midrule

\textbf{\multirow{4}{*}{GPT-4o}} &\textbf{Basic zero-shot}      & 0.67 & 0.81 & 0.75 & 0.78 \\
                                     &\textbf{Enhanced zero-shot}   & 0.67 & \textbf{0.85} & 0.70 & 0.76 \\
                                     &\textbf{Few-shot}             & 0.61 & \textbf{0.85} & 0.62 & 0.72 \\
                                     &\textbf{CoT}                  & 0.68 & 0.82 & 0.73 & 0.77 \\
                                     &\textbf{RAG}                  & \textbf{0.80} & \textbf{0.85} & \textbf{0.90} & \textbf{0.88} \\
\midrule

\textbf{\multirow{4}{*}{cross-validation}} & \textbf{zero\_shot}           & 0.70      & 0.80       & 0.83   & 0.81 \\
                                            &\textbf{enhanced\_zero\_shot} & 0.70      & 0.82      & 0.79   & 0.81 \\
                                            &\textbf{few\_shot}            & 0.73     & 0.82      & 0.83   & 0.83 \\
                                            &\textbf{Cot}                  & 0.66     & 0.83      & 0.71   & 0.77 \\
                                            &\textbf{RAG}                  & \textbf{0.80}     & \textbf{0.84}      & \textbf{0.90}   & \textbf{0.87}  \\
\midrule
\textbf{\multirow{2}{*}{temporal split}} &\textbf{RAG (all retrieval)}                 & 0.76     & 0.83      & 0.85   & 0.84 \\
                                         &\textbf{RAG (post-2023.9.1 retrieval)} & 0.75     & 0.83      & 0.84   & 0.83 \\

\bottomrule
\end{tabular}
\end{threeparttable}
}
\end{table}

\subsection{Implications}
Our findings provide actionable implications for Linux kernel maintainers, users, and researchers working on automated bug report triage. Below, we summarize the most salient recommendations:

\textbf{Kernel Maintainers.} False-positive reports impose comparable effort to genuine bugs (Finding 1), highlighting the need for automated triage support. 
Component-specific susceptibility (Findings 2-4) further suggests that component-aware validation and targeted documentation, especially for Drivers and File Systems, can reduce unnecessary debugging cycles. 
Moreover, the prevalence of dependency-related and misunderstanding-driven False-positves (Findings 5, 6) underscores the importance of clearer guidance on implicit behaviors and environment setup.

\textbf{Reporters and Fuzzing Tool Developers. } Many human-reported false positives stem from misinterpretations of component behaviors (Finding 3, 7). Users would benefit from improved submission interfaces with pre-checks, documentation search, and duplicate detection to avoid redundant reports, as well as explicit guidance to first attempt reproduction on a mainline kernel version or by varying their dependencies and hardware configurations. On the other hand, syzkaller-generated false positives are concentrated in Networking and File Systems (Finding 4, 7), exposing limitations in fuzzing design. Enhancing environment modeling, dependency validation, and behavior-aware oracle design could reduce these systematic False-positive patterns. Together, these improvements would lower the reporting burden across both human and automated channels.

\textbf{Researchers.} Findings 8, 9 highlight both the potential and the limitations of LLMs in triaging false-positive bug reports. While zero-shot prompting proves inadequate, retrieval-based prompting substantially improves accuracy by grounding classification in contextual information such as documentation or prior bug reports. In contrast, chain-of-thought prompting underperforms despite its additional reasoning steps, suggesting that factual grounding is more effective than speculative reasoning in this domain. Moreover, recall emerges as the key differentiator across prompting strategies, underscoring the need to prioritize methods that minimize missed false positives. Finally, although LLM explanations are generally plausible, they can be superficial, pointing to the value of refining explanation quality through techniques such as model ensembles, post-hoc verification, or alignment with structured root-cause taxonomies.

\section{Threats to Validity}

\textbf{External validity.} Our study is based on false-positive bug reports drawn from the Linux kernel’s Bugzilla and Syzkaller. While these two sources cover both human- and fuzzing-generated reports and thus provide a complementary view, they may not represent the full spectrum of bug reporting practices in other operating systems or projects. In particular, proprietary kernels, embedded platforms, or other fuzzing tools may exhibit different distributions of false positives and root causes. Still, Bugzilla and Syzkaller are the primary entry points for bug reporting, which gives us confidence in the relevance of our dataset.

\smallskip
\noindent
\textbf{Internal validity.} A potential threat comes from the way we annotated false-positive reports and root causes. To reduce the subjective bias, we only included closed reports that had received developer responses, which means false positives in reports without responses may not be captured. Nevertheless, some degree of subjectivity remains. To mitigate this, the labeling was performed collaboratively by multiple researchers, with any disagreements resolved through discussion.
Another source of internal threat is the evaluation of LLM-based classification. The outcomes may be influenced by prompt design choices, parameter settings, or the nondeterministic nature of model outputs. We attempted to minimize these effects by standardizing the experimental setup and running controlled comparisons across strategies, but reproducibility can still be partially affected.

\smallskip
\noindent
\textbf{Construct validity.} Our conclusions rely on how we defined and operationalized the notion of false positives and their root causes. While our taxonomy was informed by prior research and iteratively refined through pilot studies, it may not capture all subtle forms of misunderstandings or dependency-related issues. Moreover, relying on developer comments to validate whether a report is false positive may overlook cases where the reasoning was incomplete or implicit. To mitigate this, we emphasized transparency in our coding process and consistently grounded our labels in the resolution outcomes.

\section{Conclusion}
\label{sec:conclusion}

This study provides the first systematic investigation of false-positive bug reports in the Linux kernel, revealing both their prevalence and their cost to the community. Our empirical analysis shows that false positives consume substantial developer effort, rivaling that of genuine bugs, while ultimately producing no corrective patches. We further uncover that their occurrence is unevenly distributed across kernel components and largely driven by dependency misconfigurations and semantic misunderstandings. Building on these insights, we demonstrate that large language models can serve as effective, interpretable tools for reducing the triage burden. Retrieval-augmented prompting, in particular, proves capable of balancing accuracy, recall, and explanation quality, offering maintainers actionable support in distinguishing false positives from genuine issues. While our findings are specific to the Linux kernel, the methodological framework and lessons extend to other large-scale, safety-critical software ecosystems. By releasing a curated dataset and outlining practical mitigation strategies, we aim to enable both the research community and practitioners to reduce wasted engineering effort and to advance the reliability of bug reporting systems.

\section*{Data Availability} 
To facilitate future research and practical use, we release our replication package including (1) 497 Linux kernel false-positive bug reports with corresponding metadata and 1,509 genuine bug reports; and (2) scripts used for LLM-based mitigation strategies. The package is available at \url{https://github.com/tianjiashuo/False-Positive-from-Linux-Kernel}.

\begin{acks}
This work is supported by National Key Research and Development Program of China (No. 2024YFB4506300), and National Natural Science Foundation of China (Grant Nos. 62322208 and 12411530122).
\end{acks}

\normalem
\bibliographystyle{ACM-Reference-Format}
\bibliography{13_ref}

\end{document}